\documentclass[aps,prl,showpacs,twocolumn,floatfix,superscriptaddress]{revtex4-1}

\usepackage{amsmath,amssymb}
\usepackage{graphicx}
\usepackage{dcolumn}
\usepackage{bm}
\usepackage[usenames]{color}
\usepackage[breaklinks=true,colorlinks=true,linkcolor=blue,urlcolor=blue,citecolor=blue]{hyperref}
\usepackage{ulem}
\usepackage{amsmath}
\usepackage{ulem}

\usepackage{amsfonts}%
\usepackage{amsthm}%
\usepackage{mathrsfs}%
\usepackage{gensymb}%

\begin{document}

\date{\today}

\title{Molecular Dynamics Study of Electro-Osmotic Flow in a Nanochannel with Molybdenum Disulfide Walls}

\author{S.M.Kazem Manzoorolajdad}
\email{kazem.manzoor@email.kntu.ac.ir}
\affiliation{Department of Physics, K.N. Toosi University of Technology, Tehran, 15875-4416, Iran}
\affiliation{School of Nano sciences, Institute for Research in Fundamental Sciences (IPM), 19395-5531, Tehran, Iran.}

\author{Hossein Hamzehpour}
\email{hamzehpour@kntu.ac.ir}
\affiliation{Department of Physics, K.N. Toosi University of Technology, Tehran, 15875-4416, Iran}
\affiliation{School of Nano sciences, Institute for Research in Fundamental Sciences (IPM), 19395-5531, Tehran, Iran.}

\author{Jalal Sarabadani}
\email{jalal@ipm.ir}
\affiliation{School of Nano sciences, Institute for Research in Fundamental Sciences (IPM), 19395-5531, Tehran, Iran.}


\begin{abstract}
The electro-osmotic flow (EOF) in a neutral system consisting of an aqueous NaCl solution confined in a nanochannel with two parallel Molybdenum disulfide ($\textrm{MoS}_{\textrm{2}}$) walls and in the presence of an external electric field parallel to the channel walls, is investigated for the first time. The results indicate that the thickness of the Stern layer grows as the negative electric surface charge density on the nanochannel walls increases. The Stern layer becomes thinner as the salt concentration is increased. Moreover, the EOF occurs under the no-slip condition on the walls. 
In addition, by increasing the surface charge density the average of the flow velocity across the nanochannel initially grows (Debye--H$\ddot{\textrm{u}}$ckel regime) and reaches its maximum value.
Then, by further increasing the surface charge density the water flow rate decreases (intermediate regime), and gets the zero value and becomes negative (reverse flow regime) at even larger values of the surface charge densities. Comparing the results of the 
previous work wherein the channels are composed of the black phosphorene walls with those of the present study for a channel composed of $\textrm{MoS}_{\textrm{2}}$ surfaces, show that for the latter case the reverse flow occurs at a lower surface charge density and with a greater value of the peak velocity with respect to the change in the surface charge density for the former case.
\end{abstract}

\maketitle


In micro-/nanoscale systems, electrokinetics plays a crucial role in the dynamics of liquid, ions and bioparticles. It has a wide range of applications, including desalination, drug delivery, micro- and nano-pumping, separation, geoscience and energy storage~\!\cite{cohen2012water,angelova2008dynamic,ramsey1997generating,hamzehpour2014electro,merlet2014electric,%
zhang2022diffusio,hardt2020electric,faraji2020electrokinetic,chowdhury2022electrophoresis,%
zhang2009dielectrophoretic}. Multiple variables, such as surface tension gradient, temperature, pressure, electric field, and concentration gradient can affect the fluid flow within nanochannels~\!\cite{squires2005microfluidics,schoch2008transport}. Compared to other propelling forces, electro-osmotic flow (EOF) provides a rapid and efficient method for controlling the flow in highly confined environments. The EOF is the motion of the electrolyte induced by applying an external electric field to the net charge within the system's electric double layer (EDL). When quenched charges on a dielectric surface are brought into contact with an ionic solution, the surface is rapidly charged due to protonation/deprotonation reactions with the ionic solution. This results in an EDL with a net charge, which is formed by counterions \cite{karniadakis2006microflows}. The excess dissolved ions in the EDL's diffuse layer move with or against the applied electric field, dragging the surrounding water molecules. The thickness of EDL varies depending on the system under consideration, often ranging from 1~\!nm for solutions with high conductivity to 100~\!nm for deionized water\cite{dutta2002electroosmotic}.
The assumptions of classical continuum mechanics and the Poisson-Boltzmann theory may not provide a comprehensive characterization of the flow field and ion distribution in nanofluidic systems. The classic Poisson-Boltzmann formalism, which exclusively accounts for electrostatic interactions among charged species, fails to provide a comprehensive explanation for the presence of significant surface charges and the influence of ion valency. Furthermore, the original PB theory \cite{silalahi2010comparing,tresset2008generalized,fleck2005counterion,boroudjerdi2005statics} did not consider the variations in ion and water densities at the surface, as well as the discontinuous effects resulting from finite molecule diameters. The continuum models fail to incorporate the phenomenon of flow reversal and several anomalous transport phenomena that arise due to charge-inversion and ion-specific effects \cite{qiao2004charge,huang2007ion,cao2018anomalous,rezaei2015surface,rezaei2015molecular,%
liu2020temperature,telles2022reversal,greberg1998charge,bag2018electroosmotic,lin2020charge}. In some studies, classical theories are modified in order to make accurate predictions \cite{celebi2018molecular,celebi2019molecular}. In contrast, molecular dynamics (MD) simulations offer enhanced precision through the incorporation of molecule-level intricacies. Holt {\it et al.} conducted a study that revealed a significant disparity between the flow velocity observed within carbon nanotubes and the predictions made by continuous hydrodynamic models employing stick boundary conditions. The observed flow velocity was found to be more than three orders of magnitude higher than the expected values \cite{holt2006fast}.
The characteristics of EOF and EDL are significantly affected by several factors, including the induced surface charge, the surface parameters of the nanochannel walls (such as geometry and wettability), the characteristics and concentrations of ions, and the surface chemistry. The effects of surface charge on surface wettability and wetting kinetics were investigated by Puah {\it et al.}~\!\cite{puah2010influence}.

The effect of salt concentration on the Stern layer thickness studied by Brown {\it et al.} experimentally, and they found that the Stern layer becomes thinner as the electrolyte concentration is increased~\!\cite{brown2016effect}.

In recent years, researchers have had attention on two-dimensional materials and using these materials in water transport and having control on flow. Cao and Netz have investigated the effect of electroosmosis on the channels made of graphene and have shown that when the sign of the induced electric charge on the walls is positive, with the increase of the electric charge, the flow speed first decreases and then increases~\!\cite{cao2018anomalous}. In another study, the electroosmosis has been studied in phosphorene nanochannels, wherein the authors have examined the effect of the four allotropes of phosphorene, i.e. black, blue, green and red phosphorene on the characteristics of the electroosmosis~\!\cite{kazem2023electro}. One of the specific two-dimensional materials that has attracted a lot of interests is molybdenum disulfide ($\textrm{MoS}_{\textrm{2}}$). The single-layer membrane of $\textrm{MoS}_{\textrm{2}}$ has an approximate thickness of 10~\!\textrm{\AA} and Young's modulus of 270 $\pm $ 100~\!GPa~\!\cite{barati2014thermodynamic}. Moreover, $\textrm{MoS}_{\textrm{2}}$ with its outstanding features, has attracted a wide attention for its potential applications in detecting and identifying the DNA and amino acids~\!\cite{farimani2014dna,barati2018identification}. As an example, a recent experiment showed that $\textrm{MoS}_{\textrm{2}}$ sheets have good stability in water and do not react with water~\!\cite{li2019experimental}.
In addition, the high mechanical strength and non-reactive water transport capability make the $\textrm{MoS}_{\textrm{2}}$ a superior candidate for use as walls of a nanochannel. Many studies have been performed on the filtration by using $\textrm{MoS}_{\textrm{2}}$~\!\cite{heiranian2015water,cao2020single,yao2022molecular}, and as far as we are aware of there is not any study on electroosmotic flow in a nanochannel composed of $\textrm{MoS}_{\textrm{2}}$ sheets. Therefore, in the present paper our aim is to study the EOF in channels with $\textrm{MoS}_{\textrm{2}}$ walls and to examine the effect of surface charge density of the $\textrm{MoS}_{\textrm{2}}$ walls on both the velocity of electroosmosis and the Stern layer. Our results show that the surface charge density significantly affects the fluid flow behavior.

The remainder of the paper is structured as follows: In the methods section, the molecular dynamics simulations are described in detail. Then the section devoted to results that includes three subsections: EDL in the vicinity of the $\textrm{MoS}_{\textrm{2}}$ walls, EOF velocity, and EOF in channel composed of $\textrm{MoS}_{\textrm{2}}$ walls is compared with that of the channel made of black phosphorene walls, are shown. Finally the conclusion is presented.


\section{Methods}\label{sec2}
\begin{figure}[t]
		\begin{minipage}{0.5\textwidth}
		\begin{center}
			\hspace{-0.5cm}
			\includegraphics[width=1.0\textwidth]{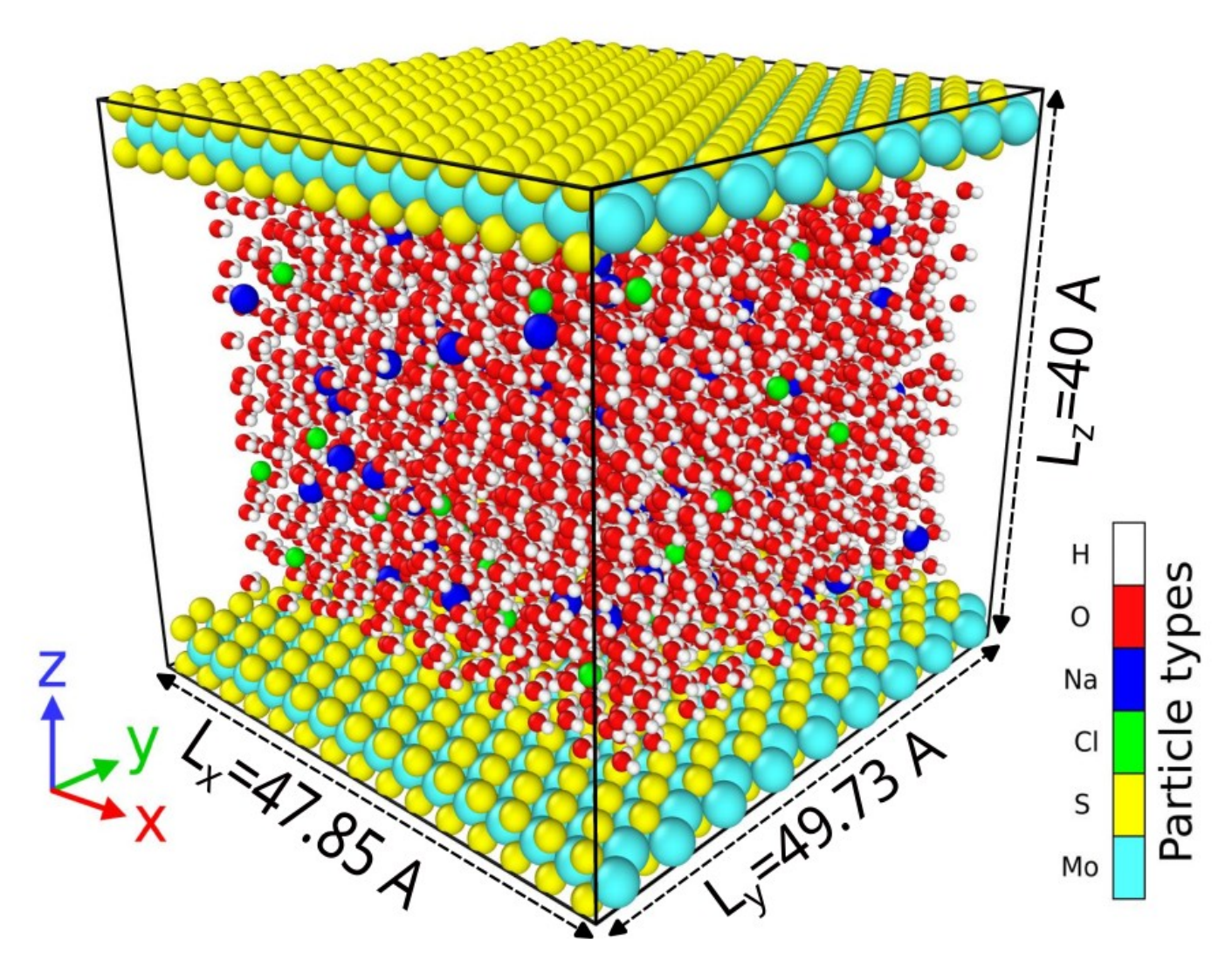}
		\end{center}
	\end{minipage}
	\caption{
A schematic representation of a neutral system consists of an aqueous NaCl solution confined in a nanochannel with two parallel $ \mathrm{MoS_2} $ surfaces. This structure has been illustrated using the OVITO package~\!\cite{stukowski2009visualization}.
	}
	\label{fig:schematic}
\end{figure}

In Fig.~\!\ref{fig:schematic} a schematic illustration of the three-dimensional model of the simulation system is shown. The system consists of two parallel $\textrm{MoS}_{\textrm{2}}$ sheets as the channel walls and the space inside the walls has been filled with an aqueous solution of sodium chloride (NaCl).
The simulation box in the $x$ and $y$ directions has dimensions of 4.785~\!nm and 4.973~\!nm, respectively, and the distance between two parallel $\textrm{MoS}_{\textrm{2}}$ walls in the $z$ direction is 4.0~\!nm. These sizes for the system have been chosen in order to prevent any overlap between the EDLs under specified conditions of surface charge density and the ionic concentration chosen in the present study.
In all of our simulations, the Mo and S atoms are placed at their original positions and are fixed, while fluid particles are allowed to move around freely within the channel. The surface charge is generated by distributing the electric point charges between Mo and S atoms. The background salt concentration are set as $C_0$=1.0~\!M and 2.0~\!M. The net negative electric charges on the walls are balanced by introducing the excess $\textrm{Na}^+$ ions in order to provide an overall charge neutrality. The ion concentration in the channel is measured using the same method as the concentration in the bulk system, given that the number of water molecules is 56 times higher than the number of NaCl molecules. 

\begin{figure}[t]
	\begin{minipage}{0.5\textwidth}
		\begin{center}
			\hspace{-0.5cm}
			\includegraphics[width=1.0\textwidth]{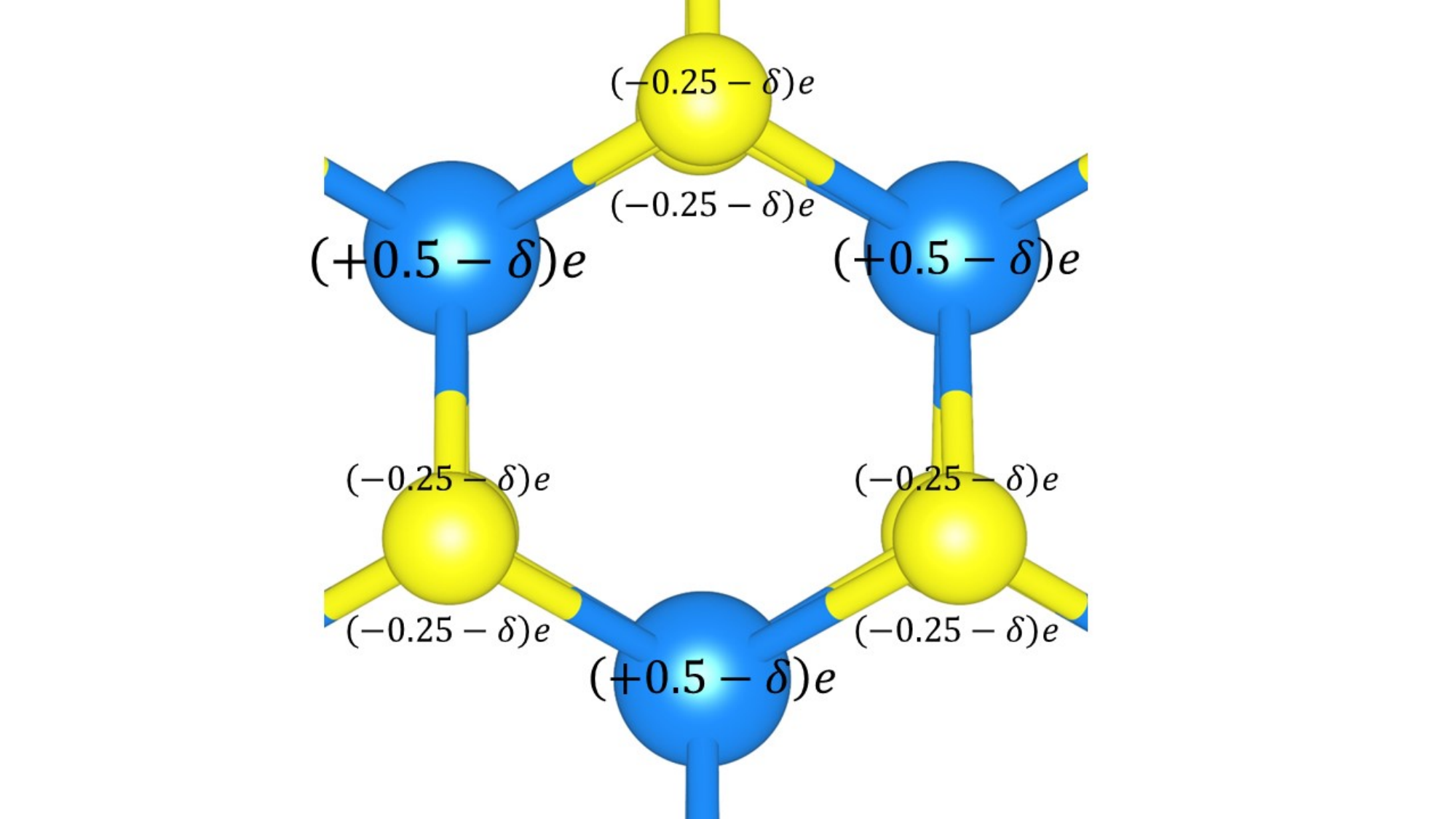}
		\end{center}
	\end{minipage}
	\caption{
		The schematic of the negative charge distribution on the atoms on the $ \mathrm{MoS_2} $ walls. 
	}
	\label{fig:charge}
\end{figure}

The schematic illustration of negative charges on $\mathrm{MoS_2}$ atoms is shown in Fig.~\!\ref{fig:charge}, in which the $\delta$ is the amount of the electric charge that is added to the atoms in the $\mathrm{MoS_2}$ sheet to make the nanochannel walls negatively charged. Therefore, the value of the charge density $\sigma$ depends on the value of the $\delta$.
As each $\mathrm{MoS_2}$ wall has surface roughness, $\sigma$ is given by
\begin{equation}
	\sigma=\frac{1}{2A_{\textrm{act}}} (N_{\textrm{counterion}}-N_{\textrm{co-ion}}) Q_{\textrm{e}} \;,
	\label{sigma_eq}
\end{equation}
where $A_{\textrm{act}}$ is the actual surface area, $Q_{\textrm{e}}$ is the electron charge, $N_{\textrm{counterion}}$ and $N_{\textrm{co-ion}}$ are the number of counterions and co--ions in the solution, respectively.

The water-water interaction is described by the extended simple point charge (SPC/E) model, which is frequently employed to represent aqueous NaCl solutions. This model demonstrates reliable accuracy 
and is extensively used 
for studying aqueous NaCl solutions and has proven to be effective in predicting the hydrodynamic properties of water in bulk under varying thermodynamic conditions~\!\cite{SPCE}.
The rigidity of water molecules is maintained by the use of the SHAKE algorithm~\!\cite{SHAKE-ryckaert1977numerical}. Previous studies~\!\cite{dang1991ion,brodholt1998molecular} have demonstrated the successful utilization of the Lennard-Jones (LJ) potential to simulate aqueous NaCl solutions with the SPC/E model for water. These simulations have effectively elucidated experimental observations. 
In line with this, in the current study the charged LJ atoms are employed to model the remaining components of the system, namely the ions in the solution and the atoms on the walls. The LJ potential, a widely used model in classical molecular simulations, describes the intermolecular interactions between every pair of particles and is given by
\begin{equation}
	\label{eq:1}
	V_{\textrm{LJ}}(r_{ij})=4\varepsilon_{ij} \bigg[ \bigg( \frac{\sigma_{ij}}{r_{ij}} \bigg)^{12} - \bigg( \frac{\sigma_{ij}}{r_{ij}} \bigg)^6 \bigg],  \enspace     r \leq r_{\textrm{c}}
\end{equation}
where the symbols $ \varepsilon_{ij} $  and $ \sigma_{ij}$ represent the depth of the potential well and the effective size of particles, respectively, and $ r_{ij} $ is the separation distance between two atoms. At a cutoff length $ r_c$ = 1.1~\!nm, the LJ potential is truncated and shifted.
The parameters of the LJ potential for the ions and $\textrm{MoS}_{\textrm{2}}$ atoms have been selected from Refs.~\!\cite{smith1994computer} and \cite{sresht2017quantitative}, respectively. The Lorentz-Berthelot combination has been employed to determine the interaction parameters between different atoms. Specifically, the values for $\varepsilon_{ij}$ and  $\sigma_{ij}$ are calculated using the formulas $ \varepsilon_{ij} = \sqrt{\varepsilon_{i}\varepsilon_{j}}$ and $\sigma_{ij} = (\sigma_i+\sigma_j)/2$ respectively. Due to the significantly smaller size and mass of hydrogen atoms compared to oxygen atoms, the impact of hydrogen atoms on the LJ interactions is assumed to be negligible~\!\cite{Wang2011non}. Therefore, the LJ interactions between water molecules have been determined by considering only the contributions of oxygen atoms. The interaction parameters for different atomic species are provided in Table~\!\ref{tbl:parameters}.

\begin{table}[]
	\centering
	\caption{Interaction parameters of atomic species}
	\label{tbl:parameters}
	\begin{tabular}{|c|c|c|c|}
		\hline
		\text{Atoms}    & $\sigma(\textrm{\AA})$ & $\varepsilon(\textrm{Kcal/mol})$ & q($\bar{\textrm{e}}$)   \\\hline
		$\textrm{H}$    & 0        & 0       & 0.424  \\\hline
		$\textrm{O}$    & 0.1533   & 3.166   & -0.848 \\\hline
		$\textrm{Na}^+$ & 0.0148   & 2.575   & 1      \\\hline
		$\textrm{Cl}^-$ & 0.106    & 4.448   & -1    \\\hline
		$\textrm{Mo}$   & 0.1156    & 4.43   & +0.50-\text{varying}       \\\hline
		$\textrm{S}$    & 0.4980    & 3.34   & -0.25-\text{varying}       \\\hline
	\end{tabular}
\end{table}

The electrostatic interaction between atoms i and j is described by the Coulomb potential
\begin{equation}
V_{\textrm{coulomb}} (r_{ij} )=\frac{1}{4\pi\varepsilon_0 }\frac{q_i q_j}{r_{ij}} ,
\end{equation}
where $ q_i $ and $ q_j $ are the charges on atoms i and j, respectively, and $\varepsilon_0$ is the vacuum permittivity.

The particle-particle-particle-mesh (PPPM) technique with a root mean accuracy of $ 10^{-4} $ is employed in order to enhance the efficiency of calculating long-range electrostatic interactions~\!\cite{hockney1989particle}. We use the Large-scale atomic/molecular massively parallel simulator (LAMMPS) package to perform all simulations~\!\cite{LAMMPS,Brown11,Brown12}. Periodic boundary conditions are implemented in both $x$ and $y$ directions, while a slab modification is employed to compute the electrostatic interaction for the reduced periodicity due to confinement in the $z$-direction~\!\cite{yeh1999ewald}.

The velocity-Verlet algorithm with an integrating time step of 2~\!femtoseconds is employed to numerically integrate the equations of motion. Before applying the external electric field, using the canonical ensemble with constant number of particles ($N$), volume ($V$) and temperature ($T$), the system is well equilibrated for a time interval of 2~\!nanoseconds (ns). The initial configuration of the system is prepered by distributing the random velocities among the ions and water molecules from a uniform distribution function at a temperature of 300~\!K. The Nose-Hoover thermostat is employed to sustain the system at a constant temperature.
By applying an external electric field to the fluid in the $x$ direction the phenomenon of electro-osmotic flow occurs, which is the generation of fluid motion. A velocity rescale method is employed to maintain a constant temperature, specifically by adjusting just the velocity components in the directions perpendicular to the flow, namely the $y$ and $z$ components. The external electric field applied in the present study is constant and its magnitude is 0.25~\!V/nm. Employing strong external  electric fields in experimental settings is challenging due to breakdown of the dielectric characteristics of the water. After equilibration, to reach the steady state for the flow the integrator is run for 16~\!ns. Then, in order to have good statistics, the simulations proceed further and the averages have been carried out for a time interval of 20~\!ns. To ensure that the density and velocity profiles are accurately represented with an appropriate level of detail, the space inside the channels is divided to around 80 slab-bins in the $z$ direction, each parallel to the $x$-$y$ plane.


\section{Results and Discussion}\label{sec3}

\subsection{Eelectric double layer in the vicinity of the $\textrm{MoS}_{\textrm{2}}$ wall}\label{subsec2}

When an electrolyte solution containing both positive and negative ions is positioned in close proximity to a charged wall, the resulting electrostatic force between the wall's surface and the ions causes that the counter ions to be absorbed to and co-ions to be repelled from the wall. This phenomenon creates a region in the vicinity of the wall where the concentration of counterions is higher compared to that of the co-ions. This specific region is named the electric double layer. The EDL is composed of two distinct layers: (a) the Stern layer, which is located adjacent to the solid surface and has a thickness about equal to the diameter of an individual ion, and (b) the diffuse layer, which is situated between the Stern layer and bulk region of the electrolyte solution. The ionic bulk of the system refers to the region wherein the concentrations of co-ions and counterions are the same. The Stern layer, which is situated in close proximity to the surface of the charged wall, is characterized by a complete absence of ions, and has the same charge as that of the wall. 
The particles in the Stern layer are completely influenced by the adsorbing force from the wall and are unable to move. On the other hand, while the diffusive layer is still exhibiting more concentration of counterions, it also contains some co-ions.
The particles located within the electric double layers are affected by two influential forces. The first one is the force which is applied on the particles in the electrolyte solution by the external electric field, and propels these particles in the field direction. 
The second force is the electric resistance force, which is the attraction force between the wall and the counterions. Although this force plays a significant role in the formation of the EDL, it also resists the movement of particles against the driving force.
\begin{figure*}[t]
	\begin{minipage}{1.0\textwidth}
		\begin{center}
			\includegraphics[width=1.0\textwidth]{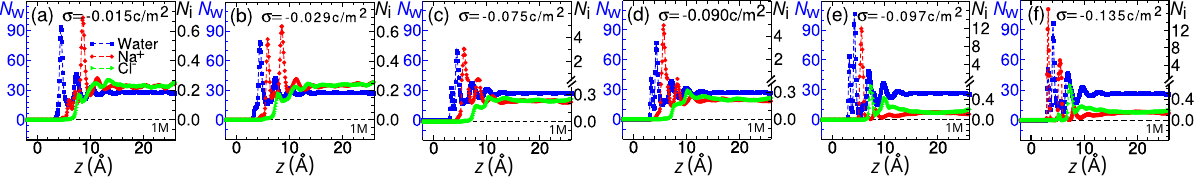}
		\end{center}
	\end{minipage} 
	\caption{
		(a) The number of water molecules (in blue color), and the number ions $\textrm{Na}^{+}$ (in red color) and $\textrm{Cl}^{-}$ (in green color) as a function of $z$ for fixed value of the surface charge density $\sigma = -0.015$~\!C/m$^2$. The external electric field is applied in the $x$ direction. Panels (b)--(f) are the same as panel (a) but for different values of the surface charge density $\sigma = -0.029$~\!C/m$^2$ to $-0.135$~\!C/m$^2$, respectively. 
	}
	\label{fig:EDL_Ex-1M}
\end{figure*}
\begin{figure*}[t]
	\begin{minipage}{1.0\textwidth}
		\begin{center}
			\includegraphics[width=1.0\textwidth]{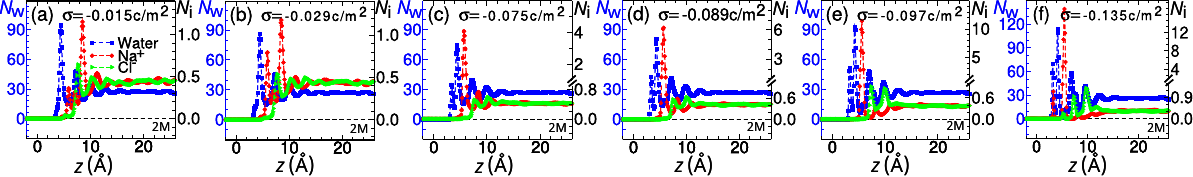}
		\end{center}
	\end{minipage} 
	\caption{
		Same as Fig.~\!\ref{fig:EDL_Ex-1M}, but for different value of the salt concentration 2.0~\!M.
	}
	\label{fig:EDL_Ex-2M}
\end{figure*}
Consequently, the net effective electrostatic force acting on each particle in the diffuse layer is weaker than that in the Stern layer. The movement of ions within the diffuse layer is influenced by an external electric field that affects the EOF. The diffuse layer is influenced by the Stern layer, that in turn leads to the regulation of the EOF. Therefore, the characteristics of the  Stern layer play important roles in the dynamics of the EOF.

It is worth noting that the Stern layer lacks a well-defined theoretical definition, and the Poisson-Boltzmann equation is able only to characterize the ion distribution in the diffuse layer.
One of the significant advantages of MD simulations in the present study is to distinguish the individual contributions of the Stern layer and the diffuse layer to the EOF. This distinction can be made by comparing the distribution profiles of co-ions and counterions in the Stern and in the diffuse layers. Hence, by employing MD simulations, a comprehensive examination of many factors influencing the EDL may be conducted.
This subsection is devoted to examination of the Stern layer, which is formed in close vicinity of the nanochannel walls. To this end the effect of different electric surface charge densities on the properties of the Stern layer are studied. 
In order to provide a quantitative description for the Stern layer, in the $z$ direction the system has been partitioned into thin rectangular cube slabs, each parallel to the $x$-$y$ plane. The sizes of each slab are $ L_x $, $ L_y $, and $\delta z = 0.1$~\!$\text{{\normalfont\AA}}$ in the $x$, $y$ and $z$ directions, respectively. To have reasonable accuracy for the desired quantities, it has been averaged over 100 different uncorrelated snapshots of the system. 
In Figure~\!\ref{fig:EDL_Ex-1M}, the average number of water molecules (blue), sodium ions (red), and chloride ions $\textrm{Cl}^-$ (green) within each thin slab are depicted as a function of the distance between the corresponding slab located at $z$ and the bottom wall of the nanochannel, for various values of the surface charge densities $\sigma=-0.015$ to $-0.135$~\!$\textrm{C}/\textrm{m}^2$ corresponding to panels (a) to (f), respectively. As seen in panels (a)-(f) by growing the absolute value of $\sigma$, the height of the first peak for the number profile of counterions $\textrm{Na}^+$ (in red color) increases, and the location of this peak is shifted to the smaller values of $z$, which means the augmented attraction towards the wall by $\textrm{Na}^+$ ions. 
Moreover, due to the excluded-volume interaction, the proximity of water molecules to the wall is reduced, that has been illustrated by plotting the distribution of water molecules in panels (a)-(f) of Figure~\!\ref{fig:EDL_Ex-1M}. As seen in panel (a), for small absolute value of $\sigma$ the peak in water number profile (in blue color) is closer to the wall than that of the $\textrm{Na}^+$ counterions  (in red color) due to the hydration of ions. As the absolute value of $\sigma$ increases from $|\sigma| = 0.015$~\!$\textrm{C}/\textrm{m}^2$ in panel (a) to $0.135$~\!$\textrm{C}/\textrm{m}^2$ in panel (f), the location of the first peak for number profile of water is shifted to the larger values of $z$, in contrast to that of the $\textrm{Na}^+$ counterions.

\begin{figure}[t]
	\begin{minipage}{0.5\textwidth}
		\begin{center}
			\hspace{-0.5cm}
			\includegraphics[width=0.75\textwidth]{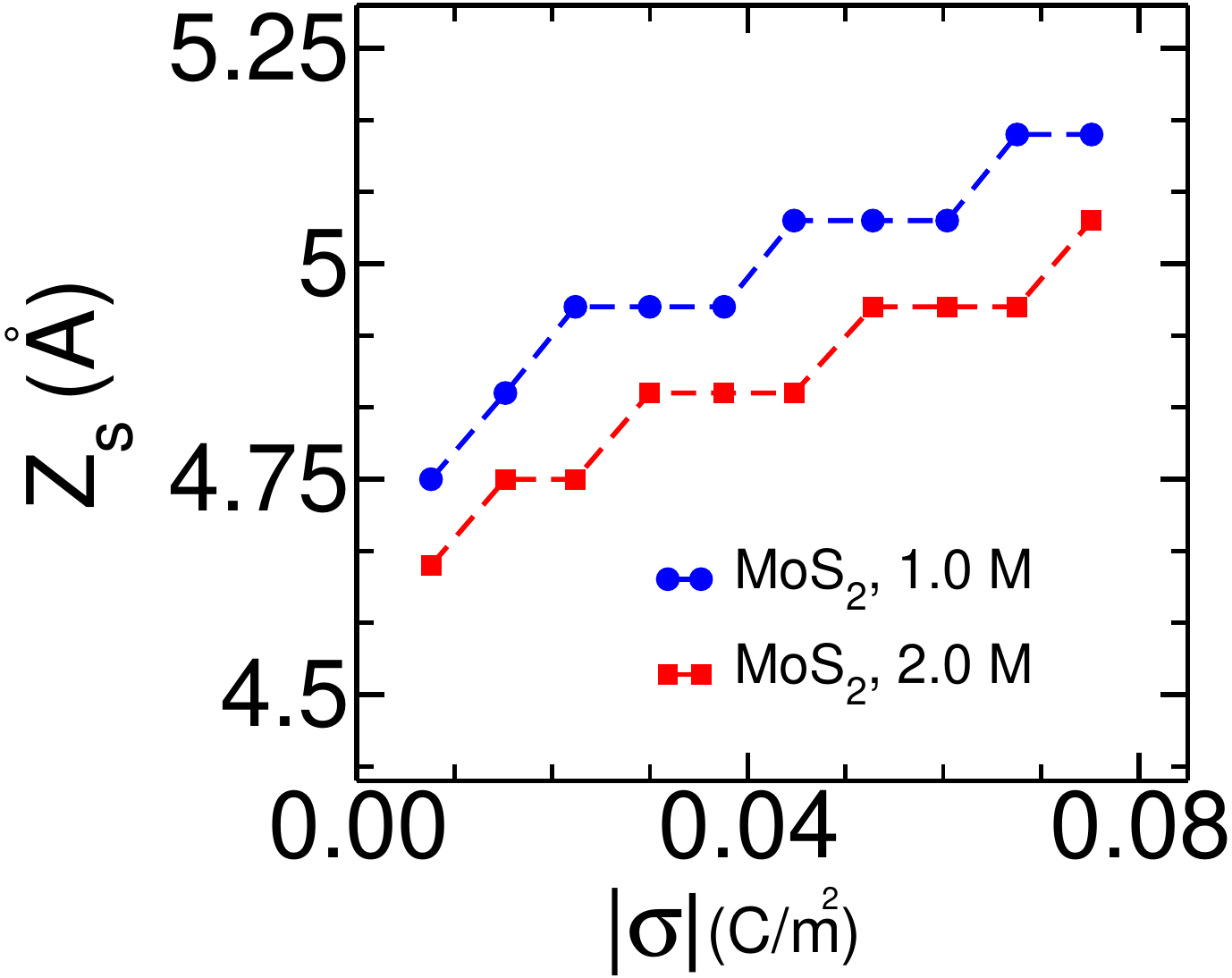}
		\end{center}
	\end{minipage} 
	\caption{
	The Stern layer thickness $Z_{\textrm{S}}$ as a function of the absolute value of the surface charge density $|\sigma|$ for 1.0~\!M (in blue color) and 2.0~\!M (in red color) salt concentrations. As the salt concentration increases, the Stern layer becomes thinner.
	}
	\label{fig:SL}
\end{figure}

To investigate the effect of ions concentration, in Fig.~\!\ref{fig:EDL_Ex-2M} the number profiles for ions and water molecules for salt concentration of $C$=2.0~\!M have been plotted as a function of $z$ for the same values of $\sigma$ chosen in Fig.~\!\ref{fig:EDL_Ex-1M}.
Though the Stern layer is not well defined in the theory, here we determine its thickness by comparing the co and counterions distribution profiles. The Stern layer is commonly described as the absorbed counterions layer close to the charged surface without any co-ions~\!\cite{wang2008electric}. In Fig.~\!\ref{fig:SL}, the thickness of the Stern layer $Z_{\textrm{S}}$ is shown as a function of the absolute value of the surface charge density for different values of the ions concentrations $C$=1.0~\!M (in blue color) and 2.0~\!M (in red color). The results show that by increasing the value of $ |\sigma| $ the value of the thickness of the Stern layer increases. The reason is that the increase in the absolute value of the surface charge density leads to more accumulation of sodium ions in this layers. Moreover, as the ions concentration increases the Stern layer becomes thinner, which is in agreement with the experimental results~\!\cite{brown2016effect,chu2023evolution,hussain2023effect}.

Finally, as seen in Figs.~\!\ref{fig:EDL_Ex-1M} and \ref{fig:EDL_Ex-2M} the size of the EDL is approximately 14~\!$\textrm{\AA}$, which is about one-third of the channel width.

\subsection{Electro-osmotic flow velocity}

\begin{figure}[t]
	\begin{minipage}{0.5\textwidth}
	\begin{center}
		\includegraphics[width=0.95\textwidth]{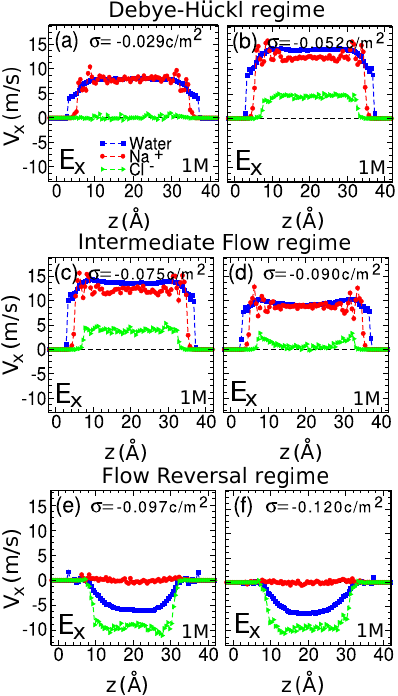}
	\end{center}
\end{minipage} 
	\caption{
(a) The velocity in the $x$ direction $v_x$ in the presence of external electric field in the $x$ direction as a function of the distance from the bottom surface $z$ for fixed value of surface charge density  $\sigma = -0.029$~\!$\textrm{C}/{\textrm{m}^2}$ for water molecules (in blue color), $\textrm{Na}^+$ counterions (in red color) and $\textrm{Cl}^-$ co-ions (in green color). Panels (b)--(f) are the same as (a) but for different values of the surface charge density $\sigma = -0.052$~\!$\textrm{C}/{\textrm{m}^2}$ to -0.120~\!$\textrm{C}/{\textrm{m}^2}$, respectively.
	}
	\label{fig:EOF-ionic}
\end{figure}

In this subsection the effect of the surface charge density $ \sigma $ on the dynamics of the electro-osmotic flow is considered by studying the velocity profiles of different species in the systems. 
To this end, extensive MD simulations have been performed for various values of the surface charge density $\sigma$ in the range of $ -0.12$~\!$\textrm{C}/\textrm{m}^2$~\!$< \sigma < -0.029 $~\!$\textrm{C}/\textrm{m}^2$.
Figure~\!\ref{fig:EOF-ionic} shows the average velocity profile for water molecules (blue line), $\textrm{Na}^+$ counterions (red line) and $\textrm{Cl}^-$ co-ions (green line) as a function of $z$ across the nanochannel for various values of the surface charge densities $-0.029$ to $-0.120$~\!C/m$^2$ corresponding to panels (a) to (f), respectively.
The physical mechanism behind the phenomenon of EOF is explained by looking at different panels in Fig.~\!\ref{fig:EOF-ionic}. Indeed, three different regimes of Debye--H$\ddot{\textrm{u}}$ckel (DH), intermediate and reverse flow are exhibited as the absolute value of the surface charge density increases.
In the DH regime, as $|\sigma|$ grows the EOF enhances, as seen in panels (a) and (b).
The present study has successfully shown that the DH regime is valid in the range $ \sigma<-0.067$~\!$\textrm{C}/\textrm{m}^2$ which corresponds to the small zeta potential in which the Poisson--Boltzmann equation can be linearized through a Taylor series expansion.
Panels (a) and (b) illustrate the velocity component $ v_x $ in the $x$ direction in the DH regime, wherein the the diffuse layer is mostly occupied by $\textrm{Na}^+$ counterions, and exhibit a predominant migration towards the direction of the externally imposed electric field. As a result, the counterions cause the transportation of the electrically neutral water molecules in the same direction as of themselves. 

The $\textrm{Cl}^-$ co-ions have tendency to travel in the opposite direction with respect to the electric field. However, in the DH regime, it is seen that co-ions experience a net force in the direction of an electric field as a result of the drag force acting on them. Consequently, their motion aligns with the electric field, but at a smaller velocity compared to the velocity of the water flow.

When the surface charge density exceeds  $-0.067$ $~\!\textrm{C}/\textrm{m}^2$, overscreening takes place, initiating the intermediate regime. In this regime, the velocity decreases as the absolute value of the surface charge density increases, but the direction of the flow is still the same as of the electric field (panels (c) and (d) of Fig.~\!\ref{fig:EOF-ionic}). 
In the intermediate regime, the presence of $\textrm{Na}^+$ counterions in the Stern layer effectively decreases the electrostatic repulsion between surface charge on the nanochannel walls and co-ions. Panels (c) and (d) illustrate a consistent drop in the velocity profile of $\textrm{Cl}^-$ co-ions. 
Furthermore, it has been shown that the velocity of $\textrm{Na}^+$ counterions decreases by increasing the surface charge density. Indeed, within the intermediate regime, the velocity profile of counterions remains above that of co-ions. Consequently, this leads to a gradual reduction in the flow of water in the direction of the electric field. 

\begin{figure}[t]
	\begin{minipage}{0.5\textwidth}
		\begin{center}
			\hspace{-0.5cm}
			\includegraphics[width=0.9\textwidth]{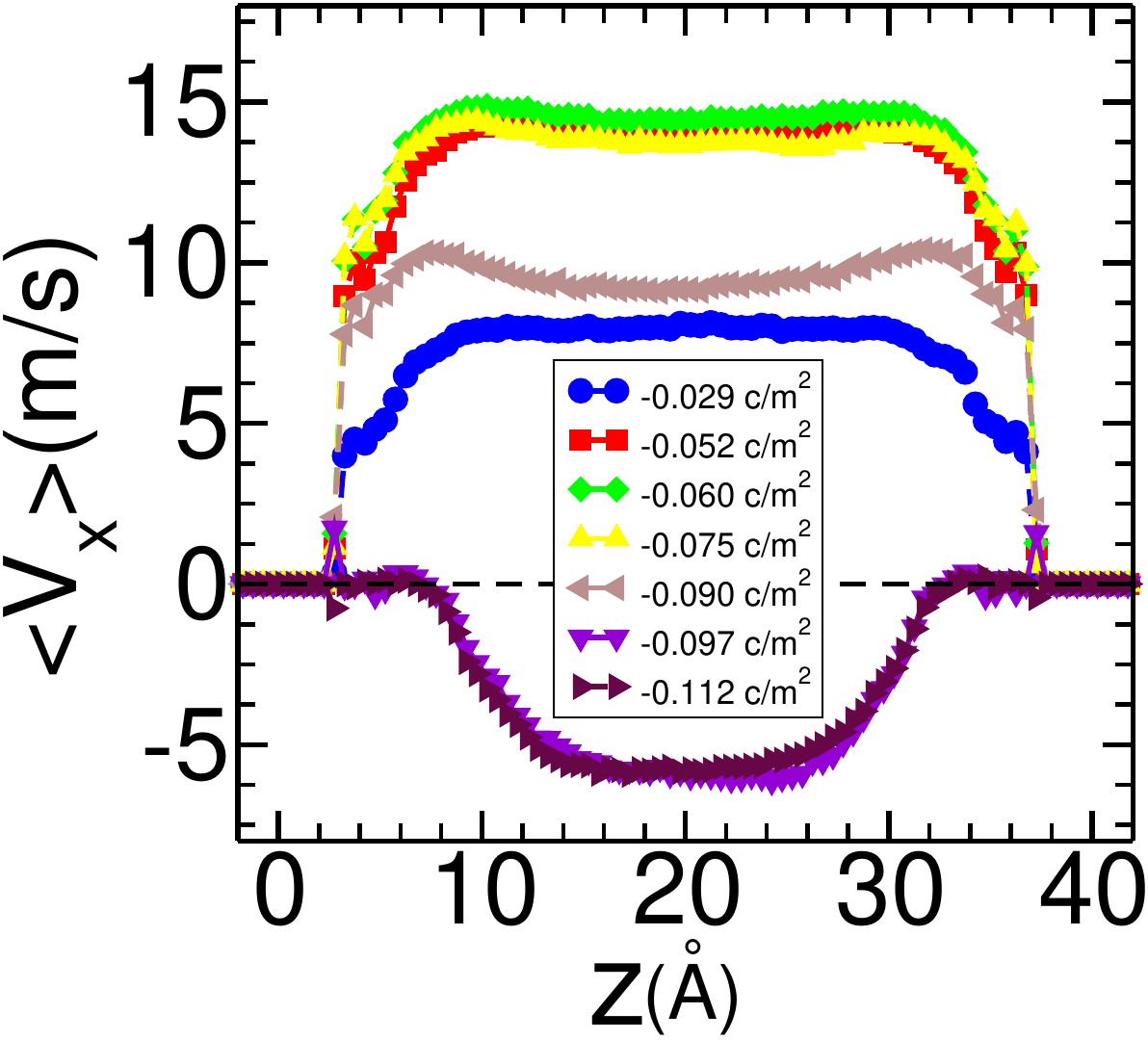}
		\end{center}
	\end{minipage} 
	\caption{
		The EOF velocity profile as a function of $z$ for different values of the surface charge densities $\sigma = -0.029$~\!$\textrm{C}/{\textrm{m}^2}$ (blue circles) to $\sigma = -0.112$~\!$\textrm{C}/{\textrm{m}^2}$ (maroon triangle right).
	}
	\label{fig:EOF-density}
\end{figure}

\begin{figure}[t]
	\begin{minipage}{0.5\textwidth}
		\begin{center}
			\hspace{-0.5cm}
			\includegraphics[width=0.8\textwidth]{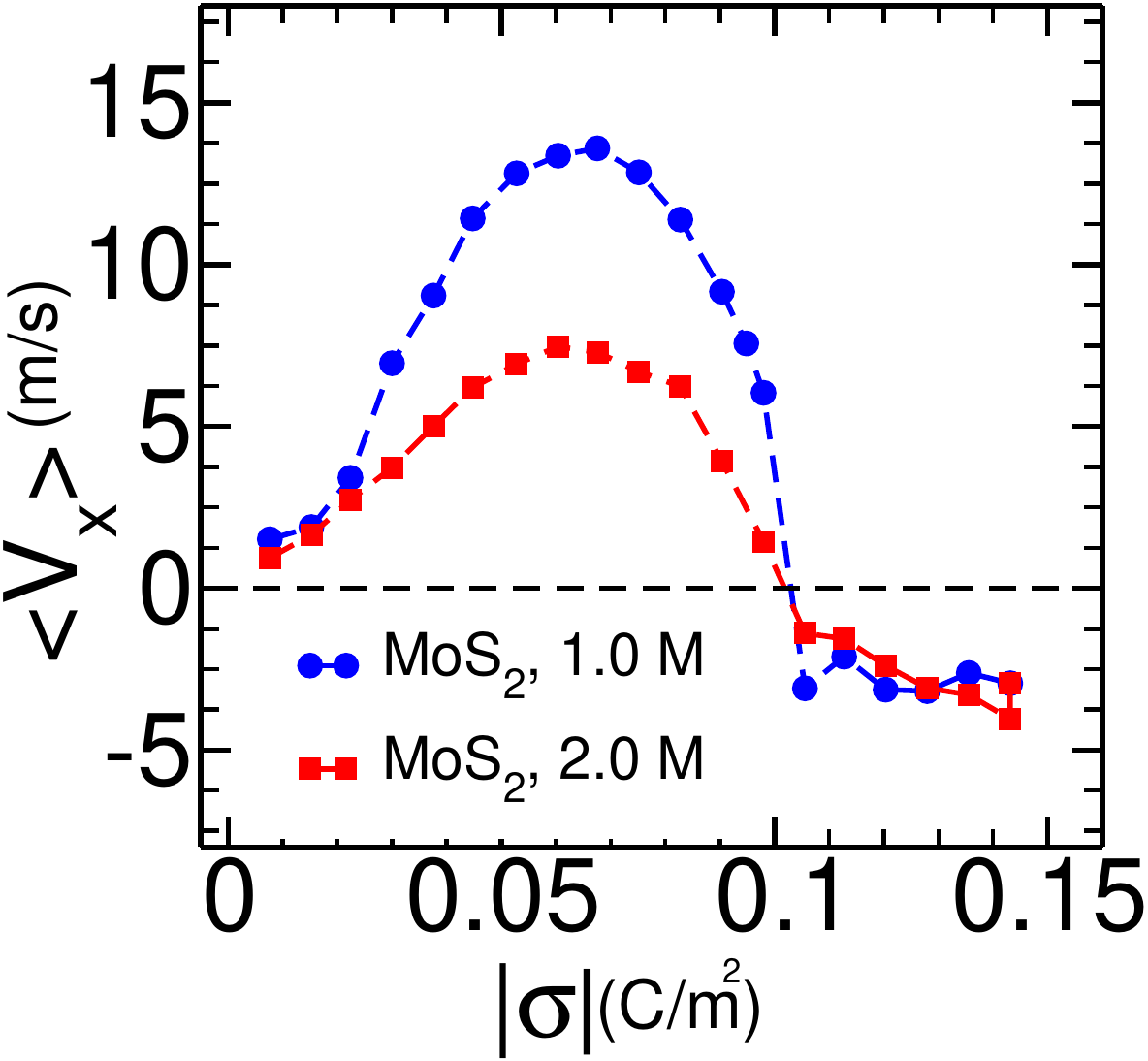}
		\end{center}
	\end{minipage} 
	\caption{
		The average flow velocities across the nanochannel as a function of the absolute value of the surface charge density $|\sigma|$ for 1.0~\!M (in blue color), 2.0~\!M (in red color) salt concentrations.
	}
	\label{fig:final}
\end{figure}

By increasing of the absolute value of the charge density even further, the velocity profile of the water molecules gets negative values, resulting in a flow reversal. The panels (e) and (f) display the flow reversal regime, wherein the flow direction is in the opposite direction with respect to the external electric field. 
Moreover, the magnitude of the average of the flow velocity increases by the increase in the absolute value of $\sigma$. 
At significantly high value of surface charge density, the strong interaction between the charged surface and the counterions leads to the predominance of the dynamics by the  $\textrm{Cl}^-$ co-ions in the diffuse layer under the influence of an external electric field, and the reverse in the flow is induced by the shear stress exerted by the mobile co-ions inside the diffuse layer. Fig.~\!\ref{fig:EOF-density} illustrates the velocity profiles of the flow as a function of $ z $ for various values of surface charge density. The average flow velocities across the nanochannel $\langle v_x \rangle$ for a certain value of the $\sigma$, is obtained by integrating the corresponding velocity profile in Fig.~\!\ref{fig:EOF-density} and then dividing by the channel width. Fig.~\!\ref{fig:final} presents $\langle v_x \rangle$ as a function of the absolute value of the surface charge density $|\sigma|$ for different values of the salt concentrations $C=1.0$ (in blue color) and 2.0~\!M (in red color), wherein the flow velocity for low salt concentration system is more than that of the high concentration.

Generally in a system consisting of surfaces with high electric charge densities, the adsorption of an excessive number of counterions in the Stern layer leads to the occurrence of overscreening, known as the charge inversion phenomenon. Indeed, the net charge of adsorbed counterions on the surface is greater than the charge of the surface itself. 
As a result, the co-ions tend to aggregate more than the counterions in the diffuse layer, which is situated between the Stern layer and the bulk of the system, that effectively modifies the polarity of the EDL. This scenario is similar to the situation in which the system consists of a wall with a positive charge. The wall comprises a negatively charged surface and a positively charged Stern layer. Additionally, there is a negatively charged diffuse layer. As a result, a secondary EDL is formed in close proximity to the positively charged sudo wall. Under the influence of the applied electric field, the co-ions associated with the recently established diffuse layer start to move in a direction opposite to that of the electric field. The phenomenon known as "reversed EOF" occurs when the water molecules are propelled in the same direction due to the movement of co-ions in the diffuse layer in the opposite direction with respect to the external field direction.
As the absolute value of the surface charge density increases beyond a certain value, e.g. $ |\sigma|>0.096$~\!$\textrm{C}/\textrm{m}^2$ for the present study, the direction of the EOF reverses to the opposite direction with respect to the direction of the external electric field.


\subsection{Comparison of EOF in channel composed of $\textrm{MoS}_{\textrm{2}}$ walls with the channel made of black phosphorene}

In this subsection the previous results for the EOF behavior of nanochannels composed of black phosphorene walls is compared with the results of the present study for the nanochannels with $ \mathrm{MoS_2} $ walls.

As Kozbial \textit {et al.}~\!\cite{kozbial2015understanding} reported water contact angle of fereshly exfoliated $ \mathrm{MoS_2} $ is 69$\pm$3.8$\degree$. Sresht \textit {et al.}~\!\cite{sresht2017quantitative} showed that the results based on the proposed force field parameter are in good agreement with those of the experiment. Moreever, Zhang \textit {et al.} reported the water contact angle on the black phosphorene sheet is 72$\degree$~\!\cite{zhang2016molecular}. In the present study for the nanochannels made of $ \mathrm{MoS_2} $ walls all simulations have been performed by employing 
parameters from Sresht's paper~\!\cite{sresht2017quantitative}, and the results for the nanochannel composed of balck phosphorene walls have been obtained by taking the parameters values from Zhang's paper~\!\cite{zhang2016molecular}.

\begin{figure}[t]
	\begin{minipage}{0.5\textwidth}
		\begin{center}
			\hspace{-0.5cm}
			\includegraphics[width=0.8\textwidth]{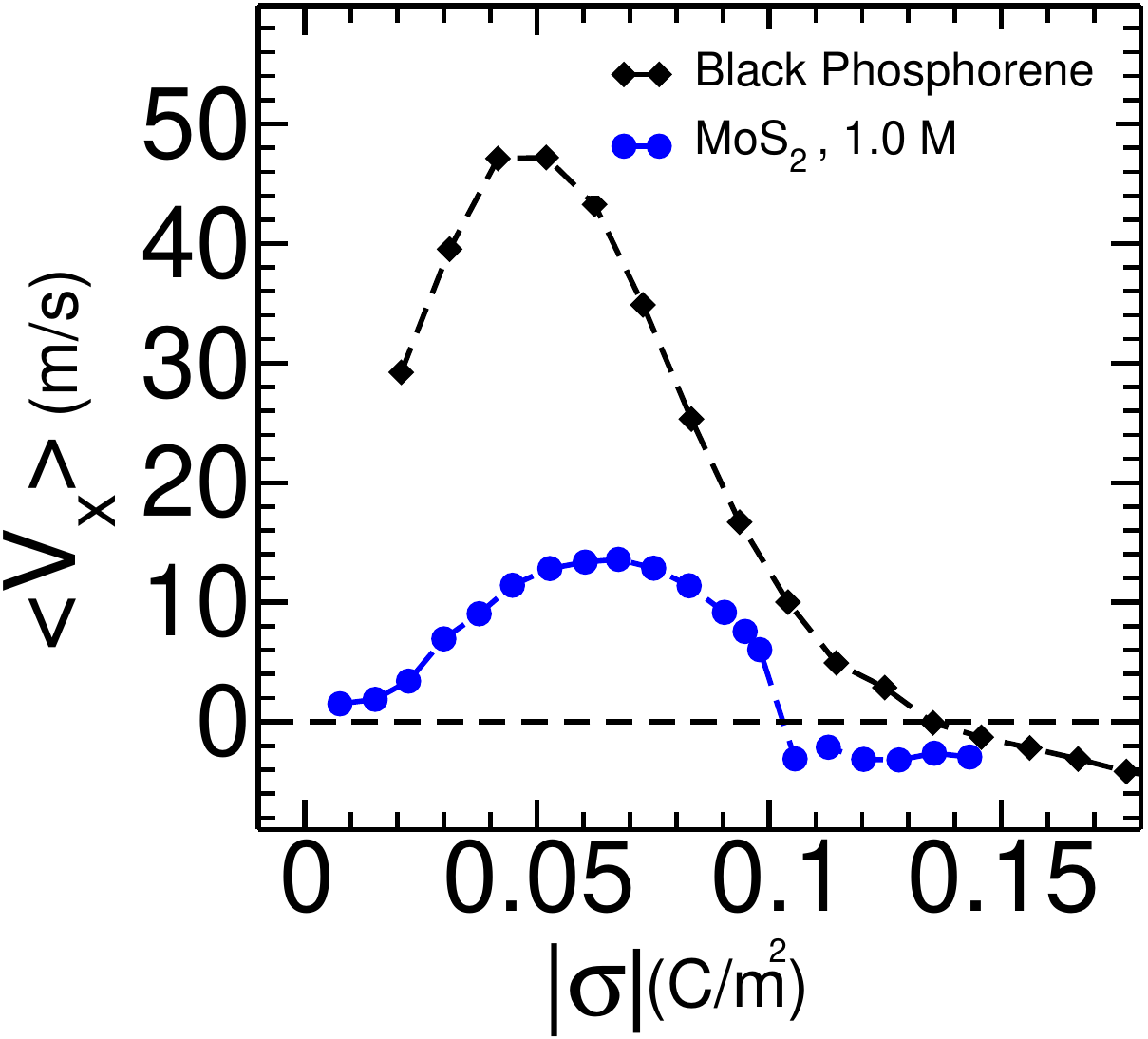}
		\end{center}
	\end{minipage} 
	\caption{
		The average flow velocity $\langle v_x \rangle $ across the nanochannel as a function of the absolute value of the  surface charge density $|\sigma|$ for a nanochannel composed of black phosphorene (in black color) and for a nanochannel made of $\textrm{MoS}_2$ sheets (in blue color). 
	}
	\label{fig:BP-MoS2}
\end{figure}

Both of black phosphorene and $ \mathrm{MoS_2} $ sheets are hydrophilic, but the latter  is more hydrophilic and experience more friction force when exposed to the water.
Therefore, the average value of the water velocity across the nanochannel made of the $ \mathrm{MoS_2} $ surfaces is smaller than that of the nanochannel composed of the Black phosphorene sheets (see Fig.~\!\ref{fig:BP-MoS2}).

Moreover, according to the results in Fig.~\!\ref{fig:BP-MoS2} for nanochannel made of Black phosphorene sheets, the absolute value of the surface charge density at which the flow reversal occurs, $|\sigma_r| $, is larger than that of the channel composed of the $\mathrm{MoS_2}$ walls. Because, as the EOF velocity for nanochannel with Black phosphorene is larger than that of the nanochannel with $\mathrm{MoS_2}$, therefore, to make the flow in the reverse direction, more electric charge density is required on the surface of the Black phosphorene walls compared on the surface of the $\mathrm{MoS_2}$ sheets.


\section{Conclusion}\label{sec13}

In summary, using the MD simulation the static and dynamic properties of the system for the EOF of an aqueous solution of NaCl between two similar parallel sheets made of the $\mathrm{MoS_2}$ have been investigated by varying the surface charge density and the salt concentration. In addition, the data showed that by increase in the absolute value of the surface charge density the thickness of the Stern layer is increased, while increase in the salt concentration leads to decrease in the thickness of the Stern layer.

%
Next, the effects of charge density and salt concentration on EOF velocity have been  examined. The results indicate that the average value of the velocity across the nanochannel increases with respect to the increase in the absolute value of the surface charge density in the DH regime, and then decreases with the increase in $ |\sigma| $ in the intermediate region. For high values of $ |\sigma| $, the EOF direction reverses with respect to the direction of the external electric field. For the system composed of the slat concentration of $C$=2.0~\!M, the average value of the velocity is smaller than that of the system with $C$=1.0~\!M. Finally, the reverse surface charge density $ \sigma_{\textrm{r}} $ (the value of $\sigma$ at which the flow reverses) for nanochannels composed of $\mathrm{MoS_2}$ walls with different salt concentrations has been investigated, and the results for the system with salt concentration of $C$=1.0~\!M has been compared to the one composed of the black phosphorene walls. The value of $ |\sigma_{\textrm{r}}| $ for the former case decreases by increasing the salt concentration,
and the comparison showed that $ |\sigma_{\textrm{r}}| $ for a channel made of black phosphorene walls is larger than that of the channel composed of $\mathrm{MoS_2}$ sheets. This happens because the $\textrm{MoS}_2$ sheet is more hydrophilic than the black phosphorene surface, and therefore the water-$\mathrm{MoS_2}$ friction is stronger than that of the water-black phosphorene. The results of our study shed light on a better understanding of the EOF dynamics inside the nanochannels made of $\mathrm{MoS_2}$ walls.




\end{document}